\documentstyle[preprint,aps]{revtex}
\input{epsf}

\def\eps{\varepsilon}

\tighten
\begin{document}
%\begin{multicols}{2}
\narrowtext
\draft
\title{Conductance length autocorrelation in quasi one--dimensional 
        disordered wires}
 
\author{Klaus Frahm\thanks{Present address: Laboratoire de Physique
 Quantique, Universit\'e Paul Sabatier, F--31062 Toulouse, France}
 and Axel M\"uller--Groeling\thanks{Present address:
 Max--Planck--Institut f\"ur Kernphysik, D--69029 Heidelberg, Germany}}
 
\address{Service de Physique de l'\'Etat condens\'e,
        CEA Saclay, 91191 Gif-sur-Yvette, France.}
 
%\date{\today}
 
\maketitle
 
\begin{abstract}
Employing techniques recently developed in the context of the
Fokker--Planck approach to electron transport in disordered 
systems we calculate the conductance length correlation function
$\langle \delta g(L) \delta g(L+\Delta L) \rangle$
for quasi 1d wires. Our result is valid for arbitrary lengths
$L$ and $\Delta L$. In the metallic limit the correlation function 
is given by a squared Lorentzian. In the localized regime it
decays exponentially in both $L$ and $\Delta L$. The correlation 
length is proportional to $L$ in the metallic regime and saturates
at a value approximately given by the localization length $\xi$
as $L\gg\xi$.

\end{abstract}
 
 
% 72.15.Rn      Quantum localization
% 05.60.+w      Transport processes: theory
% 02.50.-r      Probability theory, stochastic processes, and statistics
 
\narrowtext

%%%%%%%%%%%%%%%% Text %%%%%%%%%%%%%%%%%%%%

%%%% page 1
\section{Introduction}

In the last fifteen years two powerful approaches to calculate the
electron transport properties of quasi 1d disordered wires have been
developed: a method based on the nonlinear $\sigma$ model by Efetov
and Larkin \cite{efla} and a Fokker--Planck approach by Dorokhov
\cite{dorokhov}, Mello, Pereyra and Kumar \cite{mello2}. Both methods
are nonperturbative in the sense that they do not assume that the
system size is small compared with the localization length $\xi$.
Therefore the full range from the metallic to the localized regime can
be described within these two frameworks. Very recently it has been
shown \cite{brfr} that both approaches are in fact equivalent, despite
their considerable technical dissimilarity. Apart from the important
conceptual point of unifying our theoretical understanding of quasi 1d
wires this proof of equivalence enables us to choose among the
available techniques the more suitable one for a given problem.
Combining the results obtained within both approaches, our knowledge
about quasi 1d transport is very advanced, some would say almost
complete. We mention only a few important facts: Both the conductance
$g$ and its variance var$(g)$ have been calculated for arbitrary
system length and for all three symmetry classes \cite{mmz,brfr}.  The
probability distribution function for the generalized eigenvalues of
the transfer matrix is known exactly in the unitary case \cite{been1}
and the corresponding $n$--point correlation functions are also under
control \cite{frahm1}.  However, in spite of all this progress, there
are still a few unsolved problems.  One of the probably most important
questions concerns the magnetic field correlation function $\langle
\delta g(L,B) \delta g(L,B+\Delta B) \rangle $.  This quantity cannot
be formulated within the so--called ``minimal'' $\sigma$ model, i.e.
the supermatrix space necessary to describe this correlation function
is larger than in previous nonperturbative calculations. The technical
difficulties to treat larger $\sigma$ models exactly have so far been
insurmountable. In the Fokker--Planck approach, on the other hand, the
question is how to incorporate the magnetic field in the theoretical
formulation. This more conceptual problem also still awaits a
solution.

In this paper we calculate the conductance {\it length} correlation
function $\langle \delta g(L) \delta g(L+\Delta L) \rangle $ for all
$L$ and $\Delta L$ using the Fokker--Planck technique.  In this case
the situation is more favourable than for the magnetic field
correlation function since the system length --- unlike the magnetic
field --- is the fundamental parameter of the whole approach.
Therefore it is possible to perform an exact calculation of a quantity
that cannot be formulated within the framework of the minimal $\sigma$
model.  We briefly summarize our results: In the metallic limit
$L,\Delta L\ll\xi$ the length correlation function is given by a
squared Lorentzian, while it decays exponentially as a function of
both $L$ and $\Delta L$ in the localized regime $L,\Delta L\gg \xi$.
As $L$ increases from the metallic to the localized limit the
correlation length in $\Delta L$, which is proportional to $L$ for
$L\ll \xi$, saturates at a constant value equal to $\xi$. This
explicitly confirms the view that a strongly localized wire is
composed of independently fluctuating segments of size $\xi$.  In a
recent paper \cite{fpz}, S. Feng, J.--L. Pichard and F. Zeng have
argued that, in the localized regime, the conductance fluctuations
induced by the motion of a single impurity should be {\it universal},
$\langle [\ln g/g']^2\rangle \approx 1$. Here, $g$ and $g'$ are the
conductances before and after the impurity has moved, respectively.
The main qualitative assumption necessary to arrive at this conclusion
was precisely the decomposition into segments, which is rigorously
derived in our paper. Therefore, the universality of the conductance
fluctuations in the localized regime can now be regarded as
analytically established in quasi 1d wires. The arguments in
\cite{fpz}, however, go beyond the quasi 1d case and suggest,
supported by numerical simulations, that this universality extends
also to 2d and 3d.

%%%% NEW PARAGRAPH ABOUT A POSSIBLE EXPERIMENT
The conductance length autocorrelation function is in principle 
accessible by experiments in mesoscopic nanostructures. For this, one 
has to fabricate a ``quasi one--dimensional'' quantum--wire with several 
contacts at discrete distances $L_j$. Using such a geometry, one can measure 
the conductance as a function of $L_j$ (and of the magnetic field $B$),
where the disorder realization is {\em identical} for the common parts 
of the wire. One has to take into account, however, that in these
multi-lead devices phase coherent scattering in the attached contact
arms typically influences the experimental result.
%The mean free path and thus the localization 
%length can be varied by changing the density of impurities. 
As usual, the statistical average necessary
to calculate the mean conductance or the length autocorrelation function
may be replaced by an average over a suitable range of the 
magnetic field. 

Our paper is organized as follows. After collecting a few important
known results 
concerning the Fokker--Planck approach in section II, we develop the technical
framework for the length correlation function of an arbitrary linear statistic
in section III. In section IV we derive a (complex) expression for the
special case of the conductance length correlation function. Sections V,VI and 
VII are devoted to the metallic, localized and crossover regimes, respectively.
In section VIII we give a short summary and make some concluding remarks.
Considerations of a purely technical nature have been deferred to appendices A
to E.

\section{Known results and basic terminology}

In the Fokker--Planck approach \cite{dorokhov,mello2,mello3,mello4,macedo}
the transmission properties of a quasi 1d wire with $N$ conducting channels
and length $L$ are characterized by $N$ generalized eigenvalues 
$\lambda_1 \ge 0$ $(i=1,\ldots,N)$. They parametrize the radial part 
\cite{mello2,stone} of the transfer matrix. With $T_i=(1+\lambda_i)^{-1}$
the transmission coefficients, the conductance $g$ is given by the Landauer
formula
\begin{equation}
g = \sum_{i=1}^N T_i = \sum_{i=1}^N {1 \over 1+\lambda_i}.
\label{eq1}
\end{equation}
The multiplicative combination law for the transfer matrix yields the following
Fokker--Planck equation \cite{dorokhov,mello2,macedo} for the probability
distribution $p(\hat\lambda,t)$ of the wire,
\begin{equation}
\label{eq:1}
\partial_t p(\hat\lambda, t)=\Delta p(\hat\lambda,t).
\end{equation}
Here, $\hat{\lambda}=(\lambda_1,\ldots,\lambda_N)$ is the collection
of eigenvalues 
$\lambda_i$, $t=L/2\xi$ the length of the wire measured in units of (twice) the
localization length $\xi$, and $\Delta$ the radial part of the Laplacian on the
transfer matrix space,
\begin{eqnarray}
\label{eq:2}
\Delta & = & 4\sum_i\,\partial_{\lambda_i}\lambda_i(1+\lambda_i) 
J(\hat\lambda)\,\partial_{\lambda_i} J^{-1}(\hat\lambda)\quad,\\
\label{eq:3}
J(\hat\lambda) & = &\prod_{i>j}|\lambda_i-\lambda_j|^\beta\quad.
\end{eqnarray}
We have $\xi=(\beta N+2-\beta)l$, where $\beta=1,2,4$ corresponds to the
orthogonal, unitary, and symplectic symmetry class, respectively, and
$l$ is the 
elastic mean free path. The usual initial condition for (\ref{eq:1}) is given
by $p(\hat{\lambda},0)=\delta(\hat{\lambda})$, corresponding to ideal
transmission 
$T_i=1$ for a wire of zero length.
In the unitary case (to which we restrict our attention in this paper)
the Fokker--Planck equation (\ref{eq:1}) 
has been solved exactly \cite{been1} by means of a Sutherland--transformation.
Subsequently, the $n$--point correlation functions of the $\lambda_i$
(at a fixed 
value of the length $L$) have been calculated \cite{frahm1} for
arbitrary $N$ and 
$L$. In the following we summarize some of the results in \cite{been1,frahm1}
for further reference.

The probability distribution $p(\hat\lambda,t)$ can be expressed \cite{been1} 
in terms of a certain ``many--body'' Green's function
$G(\hat\mu,\hat\lambda;t)$  
\begin{equation}
\label{eq:4}
p(\hat\lambda,t)=\int d^N\hat\mu\ G(\hat\mu,\hat\lambda;t)
\ p(\hat\mu,0)=\lim_{\hat\mu\to 0} G(\hat\mu,\hat\lambda;t)\quad.
\end{equation}
This Green's function is given by
\begin{equation}
\label{eq:5}
G(\hat\mu,\hat\lambda;t)=\frac{1}{N!}\ \frac{\rho(\hat\lambda)}
{\rho(\hat\mu)}\ \det\left((g(\mu_i,\lambda_j;t)\right)
\,e^{C_N t}\quad,
\end{equation}
where
\begin{equation}
\label{eq:6}
\rho(\hat\lambda)=\prod_{i>j}(\lambda_i-\lambda_j)\quad,\quad
C_N=-\sum_{n=0}^{N-1} \eps_n\quad,\quad
\eps_n=(1+2n)^2\quad
\end{equation}
and
$g(\mu,\lambda;t)$ is a ``one--body'' Green's function corresponding to
%%%% page 3
the differential operator
\begin{equation}
\label{eq:7}
D(\lambda)=-(4\partial_\lambda\,\lambda(1+\lambda)\partial_\lambda+1)\quad.
\end{equation}
We have
\begin{equation}
\label{eq:8}
\partial_t g(\mu,\lambda;t)=-D(\lambda)\,g(\mu,\lambda;t)=
-D(\mu)\,g(\mu,\lambda;t)\quad,\quad 
g(\mu,\lambda;0)=\delta(\mu-\lambda)\quad.
\end{equation}
The operator $D(\lambda)$ has a continuous set of 
eigenfunctions 
\begin{equation}
\label{eq:9}
\psi_k(\lambda)=P_{\frac{1}{2}(ik-1)}(1+2\lambda)
={\textstyle F(\frac{1}{2}+i\frac{k}{2},\frac{1}{2}-i\frac{k}{2};1;
-\lambda)}
\end{equation}
with eigenvalues $k^2$, i.e. 
$D(\lambda)\psi_k(\lambda) = k^2\psi_k(\lambda)$. Here,
$P_{\frac{1}{2}(ik-1)}$ denotes the generalized 
Legendre function, which can be expressed in terms of the hypergeometric 
function $F(a,b\,;c\,;z)$  as indicated above \cite{abr}. 
The expansion of $g(\mu,\lambda;t)$ in terms of these eigenfunctions 
is given by \cite{been1}
\begin{equation}
\label{eq:10}
g(\mu,\lambda;t)=\int_0^\infty dk\ 
{\textstyle \frac{k}{2}\tanh(\frac{\pi k}{2})}\,\psi_k(\mu)\,
\psi_k(\lambda)\,e^{-k^2 t}\quad.
\end{equation}
Using (\ref{eq:5}) and (\ref{eq:10}) the limit in (\ref{eq:4}) 
can be evaluated \cite{been1}. The resulting expression can be 
rewritten \cite{frahm1} in the very useful form
\begin{equation}
\label{eq:12}
p(\hat\lambda,t)=\frac{1}{N!}\det(Q_{n-1}(\lambda_j;t))
\ \det(h_{m-1}(\lambda_i;t))\quad,
\end{equation}
where ($n,m=0,1,\ldots,N-1$)
\begin{eqnarray}
\label{eq:13}
Q_n(\lambda;t) & = & P_n(1+2\lambda)\,e^{-\eps_n t}\quad,\\
\label{eq:14}
h_m(\lambda;t) & =  & \int_0^\infty dk\ 
{\textstyle \frac{k}{2}\tanh(\frac{\pi k}{2})}\,L_m(k^2)\,
\psi_k(\lambda)\,e^{-k^2 t}\quad,\\
\label{eq:15}
L_m(z) & = & \prod_{l=0,(l\neq m)}^{N-1}\frac{z-(-\varepsilon_l)}
{(-\varepsilon_m)-(-\varepsilon_l)}\quad.
\end{eqnarray}
%%%% page 4
The Legendre polynomial $P_n(1+2\lambda)$ in (\ref{eq:13}) is also an 
eigenfunction of $D(\lambda)$. The corresponding eignevalue is
$-\eps_n$.
The $P_n(1+2\lambda)$ do not contribute in the expansion (\ref{eq:10}) 
because they are not normalizable in the range of integration $\lambda\ge 0$. 
The advantage of the representation (\ref{eq:12}) is due to the 
biorthogonality relation \cite{frahm1}
\begin{equation}
\label{eq:16}
\int_0^\infty d\lambda\ Q_n(\lambda;t)\ h_m(\lambda;t)=\delta_{nm}
\end{equation}
which was the key to calculating the $n$--point correlation 
functions \cite{frahm1}. 

In the following, we will also need the two properties
\begin{eqnarray}
\label{eq:17}
\int_0^\infty d\mu\ h_m(\mu;t)\ g(\mu,\lambda;\Delta t) & = & 
h_m(\lambda;t+\Delta t)\quad,\\
\label{eq:18}
\int_0^\infty d\lambda\ g(\mu,\lambda;\Delta t)\ Q_n(\lambda;t+\Delta t) 
& = & Q_n(\mu;t)\quad.
\end{eqnarray}
These identities can be verified using (\ref{eq:8}), (\ref{eq:14}) and the
fact that $\partial_t Q_n(\lambda;t) = D(\lambda) Q_n(\lambda;t)$ 
(see(\ref{eq:13})).

%%%% page 5
\section{Density--density correlation function for the eigenvalues}

We are finally interested in calculating the conductance length correlation
function $\langle \delta g(L) \delta g(L+\Delta L)\rangle $. To this end
we need to know the joint probability density function of finding
a certain fixed set of 
eigenvalues $\hat\mu$ at $t$ and another fixed set 
$\hat\lambda$ at $t+\Delta t$. Obviously, this density is given by
\begin{equation}
\label{eq:19}
p_2(\hat\mu,t;\hat\lambda,t+\Delta t) = p(\hat\mu;t)
\ G(\hat\mu,\hat\lambda;\Delta t)\quad,
\end{equation}
since the conditional probability density of finding a transition 
$\hat\mu\to\hat\lambda$ in a length intervall $\Delta t$ is precisely 
given by the propagator $G(\hat\mu,\hat\lambda;\Delta t)$.

We are now going to formulate a slightly more general problem than the one
we wish to solve in the end. Let 
$A_{1,2}=\sum_i a_{1,2}(\lambda_i)$ be two arbitrary linear statistics of 
the eigenvalues $\lambda_i$. Then the length correlation function of these
two quantities can be written as
\begin{equation}
\label{eq:20}
\left\langle A_1(t)\,A_2(t+\Delta t)\right\rangle=
\int_0^\infty d\lambda_1 \int_0^\infty d\mu_1
\ a_1(\lambda_1)\,a_2(\mu_1)\ R_{1,1}(\mu_1,t;\lambda_1;t+\Delta t).
\end{equation}
The function
$R_{1,1}(\mu_1,t;\lambda_1;t+\Delta t)$ is the probability density 
of finding one eigenvalue equal to $\mu_1$ at $t$ and one 
eigenvalue equal two $\lambda_1$ at $t+\Delta t$. It arises from the
joint probability density (\ref{eq:19}) upon integrating out all the
remaining eigenvalues,
\begin{equation}
\label{eq:21}
R_{1,1}(\mu_1,t;\lambda_1;t+\Delta t)=N^2\int 
d\mu_2\ldots d\mu_N\,d\lambda_2\ldots d\lambda_N
\ p_2(\hat\mu,t;\hat\lambda,t+\Delta t)\quad.
\end{equation}
Using the fact that  $\det(Q_{n-1}(\lambda_j;t)) =
const. \rho(\hat\lambda)\,e^{C_N t}$  
and with the help of 
(\ref{eq:5}), (\ref{eq:12}), and (\ref{eq:19}) we can reexpress  $p_2$ as
\begin{equation}
\label{eq:22}
p_2(\hat\mu,t;\hat\lambda,t+\Delta t) = \frac{1}{N!^2}
\ \det(h_{m-1}(\mu_i;t))\ \det(g(\mu_i,\lambda_j;\Delta t)
\ \det(Q_{n-1}(\lambda_j;t+\Delta t))\quad.
\end{equation}
%%%% page 6
Rather than calculating all the integrations in (\ref{eq:21}) explicitly, we
will employ a generating functional \cite{mehta} to calculate the density
(\ref{eq:21}): Let $u_{1,2}(\lambda)$ be arbitrary functions of the
eigenvalue $\lambda$. Following \cite{mehta} we define the functional
\begin{eqnarray}
\label{eq:23}
L[u_1,u_2] & = & \left\langle \prod_i u_1(\lambda_i(t))
\ \prod_j u_2(\lambda_j(t+\Delta t))\right\rangle\\
\nonumber
& = & \int d^N\hat\mu\int d^N\hat\lambda
\ \prod_i\Big(u_1(\mu_i)\,u_2(\lambda_i)\Big)
\ p_2(\hat\mu,t;\hat\lambda,t+\Delta t)\quad
\end{eqnarray}
so that the correlation function (\ref{eq:20}) can be expressed as
\begin{equation}
\label{eq:24}
\left\langle A_1(t)\,A_2(t+\Delta t)\right\rangle=
\partial_{z_1}\,\partial_{z_2}\,L[1+z_1 a_1,1+z_2 a_2]\Big|_{z_{1,2}=0}
\quad.
\end{equation}
Inserting (\ref{eq:22}) in (\ref{eq:23}) we find after some algebra
involving transformations of the various determinants
\begin{equation}
\label{eq:25}
L[u_1,u_2]=\det\left(M_{mn}[u_1,u_2]\right)\quad,
\end{equation}
with the $N\times N$--matrix
\begin{equation}
\label{eq:26}
M_{mn}[u_1,u_2]=\int_0^\infty d\mu\int_0^\infty d\lambda
\ h_m(\mu;t)\ g(\mu,\lambda;\Delta t)\ Q_n(\lambda;t+\Delta)
\ u_1(\mu)\,u_2(\lambda)
\end{equation}
depending linearly on $u_1$ and $u_2$. 
To calculate the derivatives in (\ref{eq:24}) we notice that due to
(\ref{eq:16})--(\ref{eq:18}) we have $M_{mn}[1,1]=\delta_{nm}$ so
that (in matrix notation)
\begin{eqnarray}
\label{eq:27}
M[1+z_1 a_1,1+z_2 a_2] &=& 1+z_1 M[a_1,1]+z_2 M[1,a_2]+z_1 z_2 
M[a_1,a_2] \nonumber\\
&\equiv& 1+X \quad.
\end{eqnarray}
To expand the determinant in (\ref{eq:25}) we employ the relation
\begin{equation}
\label{eq:28}
\det(1+X)=\exp(\mbox{tr}\,\ln(1+X))=
1+\mbox{tr}(X)+{\textstyle \frac{1}{2}}\left(
\mbox{tr}(X)^2-\mbox{tr}(X^2)\right)+\cdots\quad.
\end{equation}
Finally, using (\ref{eq:25}), (\ref{eq:27}), and (\ref{eq:28}) in
(\ref{eq:24})  we can perform the derivatives with respect to
the auxiliary variables $z_1$ and $z_2$ and arrive at
%%%% page 7
\begin{equation}
\label{eq:29}
\left\langle A_1(t)\,A_2(t+\Delta t)\right\rangle=
\mbox{tr}\Big(M[a_1,a_2]\Big)+\mbox{tr}\Big(M[a_1,1]\Big)\,
\mbox{tr}\Big(M[1,a_2]\Big)
-\mbox{tr}\Big(M[a_1,1]\,M[1,a_2]\Big)\quad.
\end{equation}
This result can now be straightforwardly compared with the expression
(\ref{eq:20}) to identify the two--point density
$R_{1,1}(\mu,t;\lambda,t+\Delta t)$. To write $R_{1,1}$ in a compact
and appropriate form we define the function
\begin{equation}
\label{eq:31}
K_N(\lambda,t_1;\mu,t_2)=\sum_{m=0}^{N-1} Q_m(\lambda;t_1)\,
h_m(\mu;t_2)\quad,
\end{equation}
in terms of which the one--point density $R_1(\mu,t)$ reads \cite{frahm1}
\begin{equation}
\label{eq:32}
R_1(\mu;t)=K_N(\mu,t;\mu,t).
\end{equation}
This quantity occurs in the calculation of simple averages of a linear
statistic such as
\begin{equation}
\label{eq:33}
\left\langle A_1(t)\right\rangle=\partial_{z_1}\,L[1+z_1 a_1,1]
\Big|_{z_1=0}=\mbox{tr}\Big(M[a_1,1]\Big)=
\int_0^\infty d\mu\ a_1(\mu)\,R_1(\mu;t)\quad.
\end{equation}
With the definitions (\ref{eq:31}) and (\ref{eq:32}) and taking into 
account the relations (\ref{eq:16})--(\ref{eq:18}) the comparison
between (\ref{eq:20}) and (\ref{eq:29}) results in
\begin{eqnarray}
\label{eq:30}
 R_{1,1}(\mu,t;\lambda;t+\Delta t) & = & R_1(\mu;t)\,
R_1(\lambda;t+\Delta t)\\
\nonumber
&& + K_N(\lambda,t+\Delta t;\mu,t)\Bigl[
g(\mu,\lambda;\Delta t)-K_N(\mu,t;\lambda,t+\Delta t)\Bigr]\quad.
\end{eqnarray}
To summarize the result of this section, the correlator of $A_1(t)$
and $A_2(t+\Delta t)$ is given by
\begin{eqnarray}
\nonumber
\left\langle \delta A_1(t)\,\delta A_2(t+\Delta t)\right\rangle & = &
\left\langle A_1(t)\,A_2(t+\Delta t)\right\rangle-
\left\langle A_1(t)\right\rangle\,
\left\langle A_2(t+\Delta t)\right\rangle\\
\label{eq:34}
& = & \int_0^\infty d\lambda \int_0^\infty d\mu
\ a_1(\lambda)\,a_2(\mu)\ S(\mu,t;\lambda;t+\Delta t) \quad,
\end{eqnarray}
where we have introduced 
the density--density correlation function for the generalized
eigenvalues of the transfer matrix,
\begin{equation}
\label{eq:35}
S(\mu,t;\lambda;t+\Delta t)=K_N(\lambda,t+\Delta t;\mu,t)\Bigl[
g(\mu,\lambda;\Delta t)-K_N(\mu,t;\lambda,t+\Delta t)\Bigr]\quad.
\end{equation}
In the next section we derive an explicit expression for the conductance length
correlation function starting from (\ref{eq:34}) and (\ref{eq:35}).

%%%% page 8
\section{Condunctance length correlation function}

The correlator $\langle \delta g(t) \delta g(t+\Delta t) \rangle $ is a special
case of (\ref{eq:34}), obtained by choosing 
$a_{1,2}(\lambda) = (1+\lambda)^{-1}$,
\begin{equation}
\label{eq:36}
\left\langle \delta g(t)\,\delta g(t+\Delta t)\right\rangle =
\int_0^\infty d\lambda \int_0^\infty d\mu
\ \frac{1}{1+\mu}\ \frac{1}{1+\lambda}
\ S(\mu,t;\lambda;t+\Delta t)\quad.
\end{equation}
The evaluation of this expression is a little involved and makes use of a
few mathematical identities, which we state as we go along. To begin with, 
inserting (\ref{eq:35}) in (\ref{eq:36}) we have
$\langle \delta g(t) \delta g(t+\Delta t) \rangle = C_1+C_2$, where
\begin{eqnarray}
C_1 &=& \phantom{-} \int_0^\infty d\lambda d\mu 
{1\over 1+\lambda} {1 \over 1+\mu}
K_N(\lambda, t+\Delta t;\mu,t) g(\mu,\lambda;\Delta t) \quad,
\label{eq2} \\
C_2 &=& - \int_0^\infty d\lambda d\mu 
{1\over 1+\lambda} {1 \over 1+\mu}
K_N(\lambda, t+\Delta t;\mu,t) K_N(\mu,t;\lambda,t+\Delta t) \quad.
\label{eq3}
\end{eqnarray}
The second contribution, $C_2$, is the easier one to calculate. Replacing
the $K_N$ in (\ref{eq3}) by the r.h.s. of (\ref{eq:31}) one realizes that
the relevant remaining integral is given by
\begin{equation}
\int d\mu {1\over 1+\mu} h_m(\mu,t) Q_{m'}(\mu,t) \quad.
\label{eq4}
\end{equation}
Using the expansion (see appendix A)
\begin{eqnarray}
\label{eq:37}
\label{EQ:37}
\frac{1}{1+\lambda}\,Q_m(1+2\lambda) &=& (-1)^m e^{-\eps_m t}\left(
\frac{1}{1+\lambda}+\sum_{l=0}^{m-1} a_{ml}\,e^{\eps_l t}Q_l(1+2\lambda)
\right)\quad, \\
\label{eq:38}
a_{ml} &=& 2(-1)^{l+1}(1+2l)\sum_{\nu=l+1}^m \frac{1}{\nu}\quad\quad\quad
(a_{ml} = 0 \; \mbox{for} \; l\ge m) \quad,
\end{eqnarray}
the abbreviation
\begin{equation}
\label{eq:42}
F_m(t)=\int_0^\infty d\mu\ \frac{1}{1+\mu}\ h_m(\mu;t)\quad,
\end{equation}
and the biorthogonality relation (\ref{eq:16}) it is not difficult to
see that
\begin{equation}
\label{eq:44}
\int_0^\infty d\mu\ \frac{1}{1+\mu}\ h_m(\mu;t)\,Q_l(\mu;t)=
(-1)^l e^{-\eps_l t}\,F_m(t)+(-1)^l a_{lm} e^{(\eps_m-\eps_l)t}
\quad,
\end{equation}
and hence
\begin{eqnarray}
\nonumber
&& C_2 = -\left(\sum_{m=0}^{N-1}(-1)^m e^{-\eps_m(t+\Delta t)} F_m(t)\right)
\left(\sum_{l=0}^{N-1}(-1)^l e^{-\eps_l t} F_l(t+\Delta t)\right)\\
&&\qquad\ - \sum_{m,l=0}^{N-1} (-1)^{m+l} a_{ml}
\left(F_m(t+\Delta t)\,e^{-\eps_l \Delta t-\eps_m t}+
F_m(t)\,e^{-\eps_m(t+ \Delta t)+\eps_l \Delta t}\right)\quad.
\label{eq:45}
\end{eqnarray}
The slightly more complicated contribution $C_1$ can be evaluated as 
follows. We again replace $K_N$ by the r.h.s. of (\ref{eq:31}) and employ
the decomposition (\ref{eq:37}), so that $C_1=C_3+C_4$ with
\begin{eqnarray}
C_3 &=& \sum_{m=0}^{N-1} (-1)^m e^{-\eps_m(t+\Delta t)} \int_0^\infty
        d\lambda d\mu
        {1\over 1+\lambda} {1\over 1+\mu}
        h_m(\mu;t) g(\mu,\lambda;\Delta t) \quad, \label{eq5} \\
C_4 &=& \sum_{m,l=0}^{N-1} (-1)^m e^{(\eps_l-\eps_m)(t+\Delta t)}
        a_{ml}
        \int_0^\infty d\mu {1\over 1+\mu} h_m(\mu;t) Q_l(\mu;t) \quad.
\label{eq6}
\end{eqnarray}
To simplify (\ref{eq6}) we have taken advantage of (\ref{eq:18}). Using the
decomposition (\ref{eq:37}) in (\ref{eq6}) and recalling (\ref{eq:16}) we get
\begin{equation}
C_4 = \sum_{m,l=0}^{N-1} (-1)^{m+l} a_{ml} 
      e^{\eps_l\Delta t - \eps_m (t+\Delta t)} F_m(t) \quad.
\label{eq7}
\end{equation}
Next, to evaluate $C_3$ we insert the relations (\ref{eq:10}) and 
(\ref{eq:14}) in (\ref{eq5}) to get
\begin{eqnarray}
C_3 &=& \sum_{m=0}^{N-1} 
         \int_0^\infty dk_1 {k_1\over 2} \tanh({\pi k_1\over 2}) L_m(k_1^2)
         (-1)^m e^{-\eps_m(t+\Delta t)} e^{-k_1^2 t} \nonumber\\
&\phantom{=}& \phantom{ \sum_{m=0}^{N-1} }
         \int_0^\infty dk_2 {k_2\over 2} \tanh({\pi k_2\over 2})
         e^{-k_2^2\Delta t} \int_0^\infty d\mu  {1\over 1+\mu}
         \psi_{k_1}(\mu) \psi_{k_2}(\mu)
         \int_0^\infty d\lambda {1\over 1+\lambda} \psi_{k_2}(\lambda) \quad.
\label{eq8}
\end{eqnarray}
For further progress we need the intgral
\begin{equation}
\label{eq:39}
\label{EQ:39}
\int_0^\infty d\lambda\,(1+\lambda)^{-\alpha}\,
\psi_k(\lambda)  =  \frac
{\Gamma(-\frac{1}{2}+\alpha-i\frac{k}{2})\ \Gamma(-\frac{1}{2}+
\alpha+i\frac{k}{2})}{\Gamma(\alpha)^2}
\end{equation}
as derived in appendix B for arbitrary complex $\alpha$ with
Re$(\alpha)\ge 1/2$ and
\begin{eqnarray}
L_m(k^2) & = & \frac{(-1)^m\,4\,(1+2m)}{k^2+\eps_m}
\ \frac{a(N,m,k)}{\Gamma(\frac{1}{2}-i\frac{k}{2})\ \Gamma(\frac{1}{2}+
i\frac{k}{2})}\quad, \nonumber \\
a(N,m,k) &= &\frac{\Gamma\left(N+\frac{1}{2}+i\frac{k}{2}\right)
\ \Gamma\left(N+\frac{1}{2}-i\frac{k}{2}\right)}{
\Gamma(N-m)\ \Gamma(N+m+1)}\quad, 
\label{eq:40}
\end{eqnarray}
which follows by a straightforward calculation from (\ref{eq:15}).
Furthermore we notice that the function $F_m(t)$ defined in
(\ref{eq:42}) can be written by virtue of (\ref{eq:14}),
(\ref{eq:39}) and (\ref{eq:40}) as 
\begin{equation}
\label{eq:43}
F_m(t)=(-1)^m 4(1+2m) \int_0^\infty dk\ 
{\textstyle \frac{k}{2}\tanh(\frac{\pi k}{2})}\,
\frac{a(N,m,k)}{k^2+\eps_m}\,e^{-k^2 t}\quad.
\end{equation}
After all these steps $\langle \delta g(t) \delta g(t+\Delta t) \rangle $
is expressed as the sum of $C_2$, $C_3$, and $C_4$ as given in
(\ref{eq:45}), (\ref{eq8}), and (\ref{eq7}), respectively, after
taking into account the relations (\ref{eq:39})--(\ref{eq:43}).
Obviously, a simplification of the notation is called for. Let us
therefore define the symbol $[\ldots]^{(k,m)}_t$ by
\begin{equation}
\label{eq:47}
\left[f(k,m)\right]_t^{(k,m)}=
\sum_{m=0}^{N-1} \int_0^\infty dk\ 
{\textstyle \frac{k}{2}\tanh(\frac{\pi k}{2})}\,
\frac{4(1+2m)\,a(N,m,k)}{k^2+\eps_m}\Big\{f(k,m)\Big\} e^{-(k^2+\eps_m) t}
\end{equation}
with $f(k,m)$ an arbitrary function of $k$ and $m$. With this notation the
conductance length correlation function is given by
\begin{eqnarray}
\label{eq:48}
\left\langle \delta g(t)\,\delta g(t+\Delta t)\right\rangle & = & 
\left[S_1(k,\Delta t)\,e^{-\eps_m \Delta t} \right]_t^{(k,m)} 
-\left[S_2(m,\Delta t)\,e^{-k^2\Delta t}
\right]_t^{(k,m)}\\
\nonumber
&& -\left[e^{-\eps_m\Delta t} \right]_t^{(k,m)} \left[e^{-k^2\Delta t} 
\right]_t^{(k,m)}\quad,
\end{eqnarray}
%%%% page 11
with
\begin{eqnarray}
S_1(k,\Delta t) & = & 
{1 \over   \Gamma({1\over2} - i{k\over2})
           \Gamma({1\over2} + i{k\over2}) }
\int_0^\infty d\mu {\psi_k(\mu) \over 1+\mu}
\int_0^\infty d\lambda {g(\mu,\lambda;\Delta t) \over 1+\lambda} 
               \nonumber \\
&=& \int_0^\infty d\tilde{k} {\tilde{k}\over 2} \tanh({\pi\tilde{k}\over 2})
{ \Gamma({1\over2} - i{\tilde{k}\over2}) 
  \Gamma({1\over2} + i{\tilde{k}\over2})    \over
  \Gamma({1\over2} - i{k\over2})
  \Gamma({1\over2} + i{k\over2}) }
 \label{eq:49}
 e^{-\tilde{k}^2 \Delta t} I(k,\tilde{k}), \\
\label{eq:49a}
S_2(m,\Delta t) & = & \sum_{l=0}^{m-1} a_{ml}\,(-1)^l\,
e^{-\eps_l\Delta t}=-2\sum_{\nu=1}^m\frac{1}{\nu}\sum_{l=0}^{\nu-1}
(1+2l)\,e^{-\eps_l\Delta t}
\end{eqnarray}
and the abbreviation
\begin{equation}
\label{eq:51}
\label{EQ:51}
I(k,\tilde k)=\int_0^\infty d\mu\ \frac{1}{1+\mu}\ 
\psi_k(\mu)\,\psi_{\tilde k}(\mu)\quad.
\end{equation}
In general, the summations and integrations in (\ref{eq:48}) have
to be carried out numerically. The most difficult numerical task
is the evaluation of (\ref{eq:51}). In appendix C we derive an
expression for $I(k,\tilde{k})$ which allows for an efficient
numerical treatment. For completeness, we mention that the average
conductance can be calculated from (\ref{eq:33}) and is given by
\cite{frahm1}
\begin{equation}
\label{eq:52}
\langle g(t)\rangle = [1]_t^{(k,m)} \quad.
\end{equation}
In the limiting cases of a metallic wire ($N\rightarrow\infty,
t\ll 1$) and of a localized wire ($t\gg 1$) further analytical
progress in the calculation of the conductance length
correlation function (\ref{eq:48}) is possible. The next two
sections deal with these two limits, respectively.

%%%% page 12
\section{Metallic regime}

In this section we show that in the metallic limit $t\ll 1$ and for
large channel numbers $N\rightarrow\infty$ the conductance length
correlation function is given by a squared Lorentzian, i.e.
\begin{equation}
\label{eq:78}
\langle \delta g(t)\,\delta g(t+\Delta t)\rangle = 
\frac{1}{15}\frac{1}{(1+\frac{\Delta t}{t})^2}+{\cal O}(t)\quad.
\end{equation}
While this result is rather simple its derivation requires considerable
effort. We begin by observing that for $N\rightarrow\infty$ the
coefficient $a(N,m,k)$ (see(\ref{eq:40})) tends to unity \cite{abr}
so that the following decomposition rule for the symbol
$[\ldots]^{(k,m)}_t$ holds:
\begin{equation}
\label{eq:53}
(-\partial_t)\left[f_1(k)\,f_2(m)\right]_t^{(k,m)}=
\left[f_1(k)\right]_t^{(k)}\,\left[f_2(m)\right]_t^{(m)} \quad.
\end{equation}
Here, $f_1(k)$ and $f_2(m)$ are arbitrary functions and we have introduced 
the independent $k$-- and $m$--``averages'':
\begin{eqnarray}
\label{eq:54}
\left[f_1(k)\right]_t^{(k)} & = & 
\int_0^\infty dk\ k\tanh({\textstyle \frac{\pi k}{2}})\,e^{-k^2 t}
\,\Bigl\{f_1(k)\Bigr\}\quad,\\
\label{eq:55}
\left[f_2(m)\right]_t^{(m)} & = & 
\sum_{m=0}^\infty 2(1+2m)\,e^{-\eps_m t}\,\Bigl\{f_2(m)\Bigr\}\quad.
\end{eqnarray}
Our derivation of (\ref{eq:78}) relies essentially on the small $t$--behavior 
of the quantities $[1]_t^{(k)}$ and $[1]_t^{(m)}$, which will shortly be
seen to become important,
\begin{eqnarray}
\label{eq:56}
\label{EQ:56}
[1]_t^{(k)} & = & \frac{1}{2t}\left(1-\frac{1}{3}\, t + 
\frac{7}{30} t^2\right)+{\cal O}(t^2)\quad,\\
\label{eq:57}
\label{EQ:57}
[1]_t^{(m)}& = & \frac{1}{2t}\left(1+\frac{1}{3}\, t + 
\frac{7}{30} t^2\right)+{\cal O}(t^2)+{\cal O}(e^{-\pi^2/(4t)})\quad.
\end{eqnarray}
%%%% page 13
These expansions are derived to all orders in $t$ in appendix D. In
(\ref{eq:57}) we have also indicated the presence of {\it non--analytical}
contributions.

Let us first consider the case $\Delta t = 0$, i.e. the case of universal
conductance fluctuations. Using (\ref{eq:8}), (\ref{eq:10}), (\ref{eq:49}),
and (\ref{eq:49a}) one can show that
$S_1(k,0)=\frac{1}{4}(1+k^2)$ and $S_2(m,0)=-m(m-1)=\frac{1}{4}(1-\eps_m)$.
Hence we have from (\ref{eq:48})
\begin{equation}
\label{eq:58}
\langle \delta g^2(t)\rangle={\textstyle \frac{1}{4}}
\left[k^2+\eps_m\right]_t^{(k,m)}-\left([1]_t^{(k,m)}\right)^2
={\textstyle \frac{1}{4}}
[1]_t^{(k)}[1]_t^{(m)}-\left([1]_t^{(k,m)}\right)^2\quad.
\end{equation}
With the help of the expansions (\ref{eq:56}), (\ref{eq:57}) and using
(\ref{eq:53}) we obtain
\begin{eqnarray}
\label{eq:59}
[1]_t^{(k)}[1]_t^{(m)} & = & \frac{1}{4t^2}+\frac{4}{45}+
{\cal O}(t^2)\quad,\\
\label{eq:60}
[1]_t^{(k,m)} & = & -\int dt\left([1]_t^{(k)}[1]_t^{(m)}\right)=
\frac{1}{4t}-\frac{4}{45}\,t+{\cal O}(t^3)\quad.
\end{eqnarray}
(We omit a formal proof that the integration constant in (\ref{eq:60}) is zero.
A nonzero integration constant contradicts the fact that 
$\langle \delta g(t)^2\rangle = {\cal O}(1)$ in the metallic regime). Inserting
(\ref{eq:59}) and (\ref{eq:60}) in (\ref{eq:58}) we find the well--known
result
\begin{equation}
\label{eq:61}
\langle \delta g^2(t)\rangle =\frac{1}{15}+{\cal O}(t^2)\quad.
\end{equation}
As a side remark we mention that we have verified up to
${\cal O}(t^{22})$ (using computer algebra and higher order terms
in (\ref{eq:56}) and (\ref{eq:57}), see appendix D) that the corrections
to $\langle \delta g(t)^2\rangle $ in (\ref{eq:61}) are even in $t$
with positive coefficients. This shows explicitly that the non--analytic 
corrections indicated in (\ref{eq:57}) are vital for the onset of localization.

Now we have to generalize our treatment to the case $\Delta t>0$. We define a
new variable $x$ by $\Delta t = tx$ and will only keep the lowest relevant
order in $t$ but all orders in $x$. The expansions of $S_1(k,\Delta t)$ and
$S_2(m,\Delta t)$ in powers of $\Delta t$ read (see appendix E)
\begin{eqnarray}
\label{eq:64}
S_1(k,\Delta t) & = & \sum_{n=0}^\infty \frac{(-\Delta t)^n}{n!}
\,(-1)^n\,r_{n+1}(-k^2)\quad,\\
\label{eq:65}
S_2(m,\Delta t) & = & \sum_{n=0}^\infty \frac{(-\Delta t)^n}{n!}
\,r_{n+1}(\eps_m)\quad,
\end{eqnarray}
where $r_{n+1}(z)$ is a polynomial of degree $n+1$ in the 
%%%% page 15
variable $z$.
In appendix E, we calculate the first three coefficients 
$a_n$, $b_n$, $c_n$ in
\begin{equation}
\label{eq:66}
r_{n+1}(z)=a_n z^{n+1}+b_n z^n + c_n z^{n-1}+\cdots
\end{equation}
with the result
\begin{eqnarray}
\nonumber
a_n & = & -\frac{1}{4(n+1)^2}\qquad (n\ge 0)\quad,\\
\nonumber\\
\label{eq:67}
\label{EQ:67}
b_n & = &
\left\{\begin{array}{lcc}
\displaystyle
\frac{(2n+1)^2}{12n(n+1)}  &\quad & (n\ge 1) \\
&&\\
\displaystyle
\frac{1}{4} &\quad & (n=0) \\
\end{array}\right.
\quad,\\
\nonumber\\
\nonumber
c_n & = & 
\left\{\begin{array}{lcc}
\displaystyle
-\frac{(2n+1)(2n-1)(12n^2-5)}{180(n+1)(n-1)} & \quad & (n\ge 2) \\
&&\\
\displaystyle
-\frac{5}{16} & \quad & (n=1) \\
&&\\
0 & \quad & (n=0) \\
\end{array}\right.
\quad.
\end{eqnarray}
From (\ref{eq:64})--(\ref{eq:66}) it is clear that we will need the
following type of moments
\begin{eqnarray}
\label{eq:62}
\left[(k^2)^n\right]_t^{(k)}=(-\partial_t)^n[1]_t^{(k)} & = &
\frac{n!}{2 t^{n+1}}-\frac{1}{6}\,\delta_{n,0}+
\frac{7}{60}(\delta_{n,0} t-\delta_{n,1})+{\cal O}(t^{2-n})\quad,\\
\label{eq:63}
\left[\eps_m^n\right]_t^{(m)}=(-\partial_t)^n[1]_t^{(k)} & = &
\frac{n!}{2 t^{n+1}}+\frac{1}{6}\,\delta_{n,0}+
\frac{7}{60}(\delta_{n,0} t-\delta_{n,1})+{\cal O}(t^{2-n})\quad,
\end{eqnarray}
which have been calculated using (\ref{eq:56}) and (\ref{eq:57}).
Combining (\ref{eq:64}) and (\ref{eq:65}) with (\ref{eq:62}) and
(\ref{eq:63}) we obtain
\begin{eqnarray}
\label{eq:68}
\left[S_1(k,\Delta t)\right]^{(k)}_t & = & 
-\frac{1}{t^2}\hat a\left(\frac{\Delta t}{t}\right)
+\frac{1}{t}\hat b\left(\frac{\Delta t}{t}\right)
-\hat c\left(\frac{\Delta t}{t}\right)
+{\cal O}(t)\quad,\\
\label{eq:69}
\left[S_2(m,\Delta t)\right]^{(m)}_t & = & 
+\frac{1}{t^2}\hat a\left(\frac{\Delta t}{t}\right)
+\frac{1}{t}\hat b\left(\frac{\Delta t}{t}\right)
+\hat c\left(\frac{\Delta t}{t}\right)
+{\cal O}(t)\quad,
\end{eqnarray}
with
\begin{eqnarray}
\label{eq:70}
\hat a(x) & = & \frac{1}{2}\sum_{n=0}^\infty (n+1)\, a_n\ (-x)^n 
=-\frac{1}{8x}\ln(1+x)\quad,\\
\label{eq:70a}
\hat b(x) & = & \frac{1}{2}\sum_{n=0}^\infty b_n\ (-x)^n 
=\frac{1}{6(1+x)}-\frac{1}{24}\left(1+\frac{1}{x}\right)\ln(1+x)
\quad,\\
\label{eq:71}
\hat c(x) & = & \frac{1}{6}\,b_0-\frac{7}{60}\,a_0+
\frac{1}{2}\sum_{n=1}^\infty \frac{1}{n}\,c_n\ (-x)^n \\
\nonumber
& = & \frac{1}{10}+\frac{7x}{45}-\frac{4x^2}{15(1+x)}
+\frac{2x^3}{15(1+x)^2}-
\left(\frac{1}{72}+\frac{7}{240x}+\frac{7x}{240}\right)\ln(1+x)\quad.
\end{eqnarray}
With the help of the decomposition rule (\ref{eq:53}) we have
\begin{eqnarray}
\nonumber
(-\partial_t)\langle \delta g(t)\,\delta g(t+\Delta t)\rangle 
& = & 
\left\{\left[S_1(k,\Delta t)\right]^{(k)}_t -
\left[e^{-k^2\Delta t}\right]^{(k,m)}_t [1]^{(k)}_t\right\}
\left[e^{-\eps_m\Delta t}\right]^{(m)}_t\\
\label{eq:72}
&&-\left\{\left[S_2(m,\Delta t)\right]^{(m)}_t +
\left[e^{-\eps_m\Delta t}\right]^{(k,m)}_t [1]^{(m)}_t\right\}
\left[e^{-k^2\Delta t}\right]^{(k)}_t\quad.
\end{eqnarray}
To proceed we need the following relations
\begin{eqnarray}
\label{eq:73}
\left[e^{-k^2\Delta t}\right]^{(k)}_t & = & 
[1]^{(k)}_{t+\Delta t} = \frac{1}{2(t+\Delta t)}-\frac{1}{6}
+\frac{7}{60}\,(t+\Delta t)+\cdots\quad,\\
\nonumber&&\\
\label{eq:74}
\left[e^{-\eps_m\Delta t}\right]^{(m)}_t & = & 
[1]^{(m)}_{t+\Delta t} = \frac{1}{2(t+\Delta t)}+\frac{1}{6}
+\frac{7}{60}\,(t+\Delta t)+\cdots\quad,\\
\nonumber&&\\
\label{eq:75}
\left[e^{-k^2\Delta t}\right]^{(k,m)}_t & = & 
-\int dt\ \left[e^{-k^2\Delta t}\right]^{(k)}_t [1]^{(m)}_t\\
\nonumber
& = & \frac{1}{4}\ln\left(1+\frac{\Delta t}{t}\right)\left(
\frac{1}{\Delta t}-\frac{1}{3}+\frac{7}{30}\,\Delta t\right)
-\frac{4}{45}(t+\alpha_1 \Delta t)+\dots\quad,\\
\nonumber&&\\
\label{eq:76}
\left[e^{-\eps_m\Delta t}\right]^{(k,m)}_t & = & 
\frac{1}{4}\ln\left(1+\frac{\Delta t}{t}\right)\left(
\frac{1}{\Delta t}+\frac{1}{3}+\frac{7}{30}\,\Delta t\right)
-\frac{4}{45}(t+\alpha_2 \Delta t)+\dots\quad.
\end{eqnarray}
Here, $\alpha_1$ and $\alpha_2$ are two undetermined constants. 
However, adding (\ref{eq:75}) and (\ref{eq:76}) and expanding
both sides for small $\Delta t$ we find the constraint
$\alpha_1+\alpha_2=1$. It turns out that our final result does not
depend on the single remaining unknown constant and we obtain from
(\ref{eq:68})--(\ref{eq:76})
\begin{equation}
\label{eq:77}
(-\partial_t)\langle \delta g(t)\,\delta g(t+\Delta t)\rangle 
\simeq -\frac{2}{15}\,\frac{\Delta t}{t^2}\,
\frac{1}{(1+\frac{\Delta t}{t})^3}=
-\frac{1}{15}\,\partial_t\,\left(
\frac{1}{(1+\frac{\Delta t}{t})^2}
\right)\quad.
\end{equation}
The result (\ref{eq:78}) follows immediately upon integrating this
last relation. In principle we still have to worry about an integration
constant that might depend on $\Delta t$. However, this constant must 
vanish for $\Delta t=0$ due to (\ref{eq:61}) and possible higher order
terms are at least ${\cal O}(\Delta t) = {\cal O}(t)$.

\section{Localized regime}

In this section we derive the fact that for $t=L/2\xi \gg 1$ (localized
regime) but arbitrary channel number $N$ the conductance length
correlation function depends exponentially on both $t$ and $\Delta t$.

In the limit $t \gg 1$ only the term with $m=0$ and the region of small $k$
contribute significantly to the expression (\ref{eq:47}). This 
observation simplifies (\ref{eq:48}) enormously. Since
$[\ldots]^{(k,m)}_t \sim e^{-t}$ we can neglect higher powers of this
symbol and with $S_2(0,\Delta t) = 0$ we get
\begin{equation}
\label{eq:79}
\langle \delta g(t)\,\delta g(t+\Delta t)\rangle \simeq 
\left[S_1(k,\Delta t)\,e^{-\eps_m\Delta t}\right]^{(k,m)}_t
\simeq \langle \delta g^2(t)\rangle\ 4S_1(0,\Delta t)\,e^{-\Delta t} 
\quad.
\end{equation}
Here, we have used the fact that $\langle \delta g(t)^2 \rangle
\approx{1\over 4}[k^2+\eps_m]^{(k,m)}_t \approx 
{1\over 4} [1]^{(k,m)}_t$ (see (\ref{eq:58})) in the localized limit.
The average $\langle \delta g(t)^2 \rangle $ can be calculated by means
of a saddle--point approximation to give
\begin{equation}
\label{eq:80}
\langle \delta g^2(t)\rangle\simeq a(N,0,0)\,\frac{\pi^{3/2}}{16}\,
t^{-3/2}\,e^{-t}\quad.
\end{equation}
For $a(N,0,0)$ we have from (\ref{eq:40})
\begin{equation}
\label{eq:81}
a(N,0,0)=\frac{\Gamma(N+\frac{1}{2})^2}{\Gamma(N)\,\Gamma(N+1)}
=\frac{\pi}{2}\,\frac{(2N)!\,(2N-1)!}{4^{2N-1}\,[(N-1)!N!]^2}\quad.
\end{equation}
The function $4S_1(0,\Delta t)$ can in general (i.e. for arbitrary 
$\Delta t$) 
only be evaluated numerically. In the limiting cases 
$\Delta t\gg 1$ and $\Delta t\ll 1$ the following
approximations can be derived
\begin{equation}
\label{eq:82}
4S_1(0,\Delta t)\simeq
\left\{\begin{array}{lcc}
1-\frac{5}{4}\,\Delta t+\frac{7}{4}\,(\Delta t)^2 +\cdots & \qquad & 
(\Delta t\ll 1) \\
&&\\
I(0,0) \frac{\pi^{3/2}}{4}\,(\Delta t)^{-3/2} & & (\Delta t\gg 1) \\
\end{array}\right.
\end{equation}
where $I(0,0)=14\zeta(3)/\pi^2\simeq 1.705$ is given by the integral 
(\ref{eq:51}) at $k=\tilde k=0$. A particular consequence of our
results is the very symmetric expression
\begin{equation}
\langle \delta g(t)\delta g(t+\Delta t)\rangle = const. 
(\Delta t)^{-3/2} t^{-3/2} e^{-\Delta t} e^{-t}
\label{eq:83}
\end{equation}
for $t,\Delta t \gg 1$.

\section{Crossover regime}

To describe the crossover behavior of the conductance length
correlation function between the metallic and the localized regime
(in both $t$ and $\Delta t$) we have to rely on a numerical evaluation
of (\ref{eq:48}). We have already mentioned earlier that the key quantity
$I(k,\tilde{k})$ (\ref{eq:51}) is difficult to calculate as it stands and 
that in appendix C a more suitable expression (as far as a numerical 
treatment is concerned) is derived.

In all the computations reported here we have set the channel number
$N$ to infinity. Let us define the quantity
\begin{equation}
K(t,\Delta t) =  
{
\langle \delta g(t) \delta g(t+\Delta t) \rangle
\over
\langle \delta g(t) \delta g(t) \rangle
}
\left( 1+{\Delta t\over t} \right)^2 \quad .
\label{eqerg1}
\end{equation}

Obviously, this is the conductance length correlation function normalized
to its value at $\Delta t = 0$ and multiplied by the inverse of the
squared Lorentzian (\ref{eq:78}) to compensate for the ``trivial''
metallic behavior. By definition, $K(t,\Delta t)$ should be a constant
equal to unity for $t\ll 1$ and not too large $\Delta t$. As $\Delta t$
increases localization effects must show up and for large $\Delta t \gg 1$
we expect an exponential decay of $K(t=const.\ll 1, \Delta t)$. For
increasing $t$, on the other hand, we expect the compensation factor
$(1+\Delta t/t)^2$ to become ineffective even for small $\Delta t$ since
$K(t=const, \Delta t)$ should decay exponentially rather than algebraically
for $t\gg 1$. In Fig. 1 we have plotted $K(t,\Delta t)$ as a function of
$\Delta t$ for $t=0.1,0.2,0.5,1.0,$ and $5.0$. 
Clearly, our expectations are borne out.
However, there is one intriguing and unforeseen feature: The curve
$K(t=0.1,\Delta t)$ bears as function of $\Delta t$ striking similarities to
the behavior of the variance $\langle\delta g(t)\delta g(t)\rangle $ of the
conductance in the unitary case. This observation suggests that (\ref{eq:78})
remains true if we replace the numerical prefactor $1/15$ by the full
function
$\langle\delta g(\Delta t)\delta g(\Delta t)\rangle $, i.e.
\begin{equation}
\langle\delta g(t)\delta g(t+\Delta t)\rangle =
{\langle\delta g(\Delta t)\delta g(\Delta t)\rangle \over
\left( 1+{\Delta t\over t} \right)^2 }
+ {\cal O}(t) \quad .
\label{eqerg2}
\end{equation}
We have not tried to prove this relation analytically and a numerical
verification turns out to be difficult since we cannot go easily to
$t$ values smaller than $0.1$. Therefore (\ref{eqerg2}) remains an interesting
speculation for the time being.

Next, to characterize the distance over which two conductances are
correlated, we have determined the point $(\Delta t)_{1/e}$ such that
$\langle\delta g(t)\delta g(t+(\Delta t)_{1/e})\rangle = \langle\delta
g(t)\delta g(t)\rangle/e$. In Fig. 2 we have plotted $(\Delta
t)_{1/e}$ as a function of $t$. We see that the curve starts with a
linear behavior $(\Delta t)_{1/e} \sim t$ for $t<1$ and then saturates
at $(\Delta t)_{1/e}\approx 0.5$ for $t\gg 1$. This is exactly what we
expect from the limiting cases worked out in sections V and VI. In the
metallic regime the decay scale is set by the system length itself,
see (\ref{eq:78}). This is also dictated by the universality of the
conductance fluctuations: There can be no intrinsic length scale other
than the system length. In the localized regime, however, we have the
asymptotic expression (\ref{eq:83}) and the situation changes. The
exponential function, of course, does decay on a typical intrinsic
scale. Since $\Delta t = \Delta L/2\xi$ and the curve in Fig. 2
saturates at $(\Delta t)_{1/e}\approx 0.5$ we see explicitly that the
wire is divided into independent segments of size $\xi$.

\section{Summary and conclusions}

Employing techniques that have recently been developed to calculate
the transport 
properties of a quasi 1d wire in the Fokker--Planck approach we have
derived the 
expression (\ref{eq:48}) for the conductance length correlation function
$\langle\delta g(t)\delta g(t+\Delta t)\rangle $. In the metallic regime
($t\ll 1,\Delta t<1, N\to\infty$) we found a squared Lorentzian
$$
\langle \delta g(t)\,\delta g(t+\Delta t)\rangle = 
\frac{1}{15}\frac{1}{(1+\frac{\Delta t}{t})^2}+{\cal O}(t)\quad,
\eqno(\ref{eq:78})
$$
while in the localized regime ($t,\Delta t \gg 1$) the correlation function is
dominated by exponential tails,
$$
\langle \delta g(t)\delta g(t+\Delta t)\rangle = const. 
(\Delta t)^{-3/2} t^{-3/2} e^{-\Delta t} e^{-t} \quad .
\eqno(\ref{eq:83})
$$

For intermediate values of $t$ and $\Delta t$ we had to restrict
ourselves to a numerical evaluation of the various sums and integrals
in (\ref{eq:48}). Figs. 1 and 2 show the crossover from the metallic
to the localized regime both in the typical dependence on $\Delta t$
and in the dependence of the correlation width $(\Delta t)_{1/e}$ on
$t$. The latter quantity is proportional to $t$ in the metallic regime
and saturates at $(\Delta t)_{1/e}\approx 0.5$ (i.e.  $\Delta L
\approx \xi$) in the localized regime.

Eq.(\ref{eq:78}) generalizes the celebrated universality of the
conductance fluctuations to the case of the length correlation
function. As long as the ratio $\Delta t/t$ is constant the absolute
length of the system does not matter.  The result also shows that any
rearrangement in the wire (of the disorder potential, say) affects the
full system since the correlation width is given by the system length.
Fig. 2 proves that this changes drastically in the localized case. As
the system size exceeds the localization length $\xi$, the correlation
length no longer grows with $t$ but saturates at $\Delta L \approx
\xi$.  This demonstrates explicitly that the wire is decomposed into
statistically independent segments of size $\xi$. As a consequence,
moving one impurity no longer changes the statistical properties of
the whole wire but affects only the relevant segment.  From this
result one can deduce \cite{fpz} that the conductance fluctuations
$\langle [\ln g/g']^2 \rangle$ in a quasi 1d wire in the localized
regime are universal and of order unity.

{\bf Acknowledgements}. We thank Jean--Louis Pichard for several
helpful remarks. K.F. acknowledges C.W.J. Beenakker for fruitful
discussions. 
This work was supported by the DFG and the European
HCM program (K.F.), and a NATO fellowship through the DAAD (A.M.G.).

\begin{figure}
\epsfxsize=3.5in
\centerline{
\epsfbox{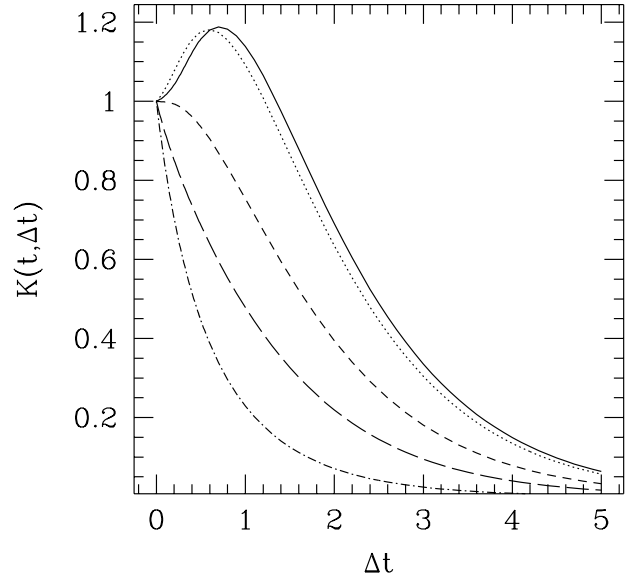}
}
\caption{
The function $K(t,\Delta t)$ as defined in the text versus $\Delta
t$. The curves correspond from top to bottom to $t=0.1$ (solid),
$t=0.2$ (dotted), $t=0.5$ (dashed), $t=1.0$ (long dashes), and $t=5.0$
(dash--dotted).
}
\label{fig1}
\end{figure}

\begin{figure}
\epsfxsize=3.5in
\centerline{
\epsfbox{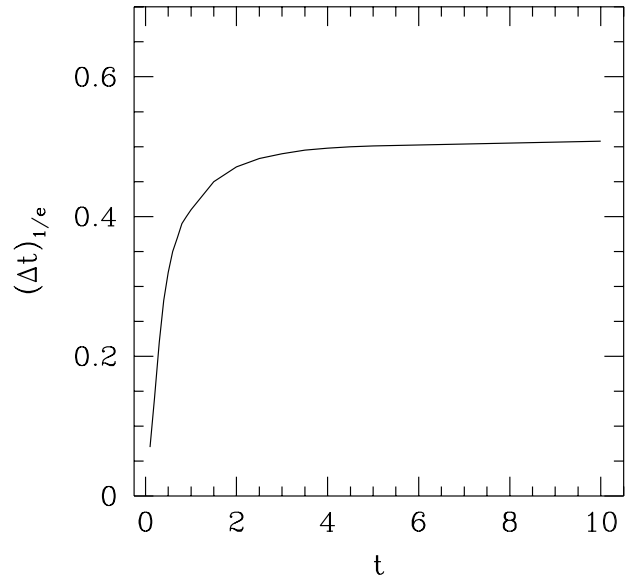}
}
\caption{
The decay width $(\Delta t)_{1/e}$ of the length correlation function
versus the system length $t$.
}
\label{fig2}
\end{figure}

%%%%%%%%%%%% Appendices
\appendix

\section{Proof of (\protect{\ref{eq:37}})} 

From the definition (\ref{eq:13}) it is clear that we have to
demonstrate that
\begin{equation}
{1\over 1+\lambda} P_m(1+2\lambda) = 
(-1)^m \left( {1\over 1+\lambda} +
        \sum_{l=0}^{m-1} a_{ml} P_l(1+2\lambda) \right)
\label{eqa1}
\end{equation}
in order to prove (\ref{eq:37}). Since $P_m(1+2\lambda)$ is a
polynomial of degree $m$ in the variable $1+\lambda$ an expansion
of the form (\ref{eqa1}) must exist. The coefficients $a_{ml}$
can be calculated as follows. With the help of the orthogonality
relation for the Legendre polynomials,
$\int_{-1}^1 dx P_n(x)\,P_m(x)=\delta_{nm}\,2/(1+2n)$, 
we get for the $a_{ml}$ ($x$ corresponds to 
$1-2(1+\lambda)$)
\begin{equation}
\label{eq:a1}
a_{ml}=(-1)^l(1+2l)\int_{-1}^1 dx\ \frac{1}{1-x}(P_m(x)-1)\,P_l(x)\quad.
\end{equation}
This integral vanishes for $m\le l$. For $m>l$ and using
$(P_m(x)-1)P_l(x)=
(P_m(x)-1)-(P_l(x)-1)+P_m(x)(P_l(x)-1)$ we find
\begin{equation}
\label{eq:a2}
a_{m}=(-1)^l (1+2l)(b_m-b_l)
\end{equation}
with
\begin{equation}
\label{eq:a3}
b_m=\int_{-1}^1 dx\ \frac{P_m(x)-1}{1-x} \quad.
\end{equation}
Applying the recurrence relation for the Legendre polynomials, 
$(2m+1)xP_m(x)=$ \\ $(m+1)P_{m+1}(x)+mP_{m-1}(x)$, 
we obtain the equation
\begin{equation}
\label{eq:a4}
b_{m+1}-b_m=-2\,\delta_{m,0}+\frac{m}{m+1}(b_m-b_{m-1})
\end{equation}
which has the solution
\begin{equation}
\label{eq:a5}
b_m=-2\sum_{\nu=1}^m\frac{1}{\nu}\quad.
\end{equation}
Together with (\ref{eq:a1}) this gives precisely the value for
$a_{ml}$ claimed in (\ref{eq:37}).

\section{The integral (\protect{\ref{eq:39}})}

In this appendix, we calculate the integral (\ref{eq:39}) for 
arbitrary complex values of $\alpha$ with Re$(\alpha)>\frac{1}{2}$. 
Using (\ref{eq:9}) and  formula 15.3.4 of Ref. \cite{abr} we get
\begin{equation}
\label{eq:b1}
{\textstyle
\psi_k(\lambda)=F(\frac{1}{2}-i\frac{k}{2},\frac{1}{2}+i\frac{k}{2};1;
-\lambda)=(1+\lambda)^{(-1+ik)/2} F(\frac{1}{2}-i\frac{k}{2},
\frac{1}{2}-i\frac{k}{2};1;\frac{\lambda}{1+\lambda})\quad.
}
\end{equation}
Substituting $s=(1+\lambda)^{-1}$ and expanding the hypergeometric 
series the l.h.s. of (\ref{eq:39}) becomes
\begin{equation}
\label{eq:b2}
\sum_{n=0}^\infty \frac{(\frac{1}{2}-i\frac{k}{2})^2_n}{n!^2}
\int_0^1 ds\ s^{-3/2-ik/2+\alpha}(1-s)^n=
\sum_{n=0}^\infty \frac{(\frac{1}{2}-i\frac{k}{2})^2_n}{n!^2}
\ \frac{\Gamma(-\frac{1}{2}-i\frac{k}{2}+\alpha)\,\Gamma(n+1)}
{\Gamma(\frac{1}{2}-i\frac{k}{2}+\alpha+n)}\quad.
\end{equation}
Here, $(a)_n=a(a+1)\cdots(a+n-1)=\Gamma(a+n)/\Gamma(a)$ is the 
Pochhammer symbol \cite{abr}. 
The evaluation of the integral involves the $\beta$-function \cite{abr}. 
The sum in (\ref{eq:b2}) can 
be expressed as a certain hypergeometric function at the special 
value $z=1$ and we can write for
the r.h.s. of (\ref{eq:b2}):
\begin{equation}
\label{eq:b3}
\frac{\Gamma(-\frac{1}{2}-i\frac{k}{2}+\alpha)}
{\Gamma(\frac{1}{2}-i\frac{k}{2}+\alpha)}\sum_{n=0}^\infty
\frac{(\frac{1}{2}-i\frac{k}{2})_n^2}{n!\,(\frac{1}{2}-i\frac{k}{2}+\alpha)_n}=
\frac{\Gamma(-\frac{1}{2}-i\frac{k}{2}+\alpha)}
{\Gamma(\frac{1}{2}-i\frac{k}{2}+\alpha)}
{\textstyle F(\frac{1}{2}-i\frac{k}{2},\frac{1}{2}-i\frac{k}{2};
\frac{1}{2}-i\frac{k}{2}+\alpha;1)}\quad.
\end{equation}
Using formula 15.1.20 of Ref. \cite{abr} we finally obtain the r.h.s. of 
Eq. (\ref{eq:39}). 

%%%% page 20

\section{The integral (\ref{eq:51})}

We derive an expression for $I(k,\tilde k)$ defined in  
(\ref{eq:51}) that is suitable for a numerical evaluation. 
Applying the transformation formula 15.3.8 of Ref. \cite{abr} 
we have
\begin{equation}
\label{eq:c1}
\psi_k(\mu)=\frac{\Gamma(ik)}{\Gamma(\frac{1}{2}+i\frac{k}{2})^2}
\,(1+\mu)^{(-1+ik)/2}\ F({\textstyle \frac{1}{2}-i\frac{k}{2},
\frac{1}{2}-i\frac{k}{2};1-ik;\frac{1}{1+\mu}})+
(k\leftrightarrow -k)\quad.
\end{equation}
We expand the hypergeometric function in (\ref{eq:c1}) and replace the first  
$\psi_k(\mu)$ in (\ref{eq:51}) by the resulting sum. For each 
term in the sum the $\mu$--integration can be done by using the 
integral (\ref{eq:39}) derived in the previous appendix. 
After some simplification the result for $I(k,\tilde{k})$ reads:
\begin{eqnarray}
\label{eq:c2}
I(k,\tilde k) & = & 
\frac{\Gamma(ik)\,\Gamma(1-\frac{i}{2}(k+\tilde k))\,
\Gamma(1-\frac{i}{2}(k-\tilde k))}
{\Gamma(\frac{1}{2}+i\frac{k}{2})^2\,\Gamma(\frac{1}{2}-i\frac{k}{2})^2}\\
\nonumber\\
\nonumber
&&\times\sum_{n=0}^\infty \frac{(1-\frac{i}{2}(k+\tilde k))_n 
(1-\frac{i}{2}(k-\tilde k))_n}{n!\,(1-ik)_n}
\ \frac{1}{(\frac{1}{2}-i\frac{k}{2}+n)^2}
 +(k\leftrightarrow -k)\quad.
\end{eqnarray}
We cannot proceed along the lines of appendix B to evaluate the sum
in (\ref{eq:c2}) due to the presence of the extra factor
$(z+n)^{-2}$ (with $z=\frac{1}{2}-i\frac{k}{2}$). A direct numerical
summation, on the other hand, is inefficient since the sum converges
only like $n^{-2}$. However, the convergence can be substantially
improved by the expansion 
\begin{equation}
\label{eq:c3}
\frac{1}{(z+n)^2}=\sum_{\nu=2}^m\frac{C_\nu(z)}{(2z+n)_\nu}+
\frac{A_m(z)}{(z+n)^2\,(2z+n)_{m-1}}+
\frac{B_m(z)}{(z+n)^2\,(2z+n)_{m}} \quad.
\end{equation}
The coefficients $C_\nu(z)$, $A_m(z)$, $B_m(z)$ depend on $z$ 
but not on $n$ and can be determined by a suitable recurrence 
relation. The contributions involving $C_\nu(z)$ can be summed analytically 
by identifying (as in appendix B)
%%%% page 21
a hypergeometric function $F(.,.;.;1)$. The remaining 
terms with $A_m(z)$ ($B_m(z)$) behave as $\sim n^{-(1+m)}$ 
($\sim n^{-(2+m)}$). Choosing $m=10-20$ the corresponding sums 
converge rapidly. 

\section{Expansion of (\ref{EQ:56}) and (\ref{EQ:57})}

This appendix deals with the expansion (in powers of $t$) of
\begin{eqnarray}
\label{eq:d1}
[1]_t^{(k)}&=&\int_0^\infty dk\ k\tanh({\textstyle \frac{\pi k}{2}})
\,e^{-k^2 t}\quad,\\
\label{eq:d2}
[1]_t^{(m)}&=&\sum_0^\infty 2(1+2m)\,e^{-(1+2m)^2\,t}\quad,
\end{eqnarray}
in the limit $t\ll 1$. Expanding $\tanh({\textstyle \frac{\pi k}{2}})$ in 
terms of exponential functions we obtain:
\begin{eqnarray}
[1]_t^{(k)} & = & \int_0^\infty dk\ k\left(1-2\sum_{n=0}^\infty(-1)^n\,
e^{-(1+n)\pi k}\right)\,e^{-k^2 t}
\nonumber\\
&&\nonumber\\
& = & \frac{1}{2t}-2\sum_{n=0}^\infty (-1)^n\int_0^\infty dk\ k 
\,e^{-(1+n)\pi k}\sum_{m=0}^\infty \frac{(-1)^m t^m}{m!}\, k^{2m}
\nonumber\\
&&\nonumber\\
& = & \frac{1}{2t}-2\sum_{m=0}^\infty \frac{(-1)^m t^m}{m!}
\,\frac{(1+2m)!}{\pi^{2(m+1)}}\sum_{n=0}^\infty (-1)^n
\frac{1}{(n+1)^{2(m+1)}}
\nonumber\\
&&\nonumber\\
& = & \frac{1}{2t}-2\sum_{m=0}^\infty \frac{(-1)^m t^m}{m!}
\,\frac{(1+2m)!}{\pi^{2(m+1)}}\left(1-\frac{1}{2^{2m+1}}\right)
\zeta(2m+2)
\nonumber\\
&&\nonumber\\
& = & \frac{1}{2t}\left(1+\sum_{m=1}^\infty \left[
\frac{(-1)^m 2(2^{2m-1}-1)}{m!}|B_{2m}|\right] t^m\right)\quad.
\label{eq:d3}
\end{eqnarray}
In the last step we have used that 
$\zeta(2m)=(2\pi)^{2m}|B_{2m}|/[2(2m)!]$, where the $B_{2m}$ denote the 
Bernoulli numbers \cite{abr}. 
%%%% page 22
With $B_2=\frac{1}{6}$ and $B_4=-\frac{1}{30}$ we find the first three 
terms in (\ref{eq:56}). 

In order to treat the discrete sum (\ref{eq:d2}) we decouple the quadratic 
term in the exponent by a Gaussian integral:
\begin{eqnarray}
[1]_t^{(m)} & = & \lim_{\eta\to 0+}\frac{1}{\sqrt{\pi t}}\,
\int_{-\infty}^\infty ds\ e^{-s^2/t}\sum_{m=0}^\infty
2(1+2m)\,e^{(1+2m)(2is-\eta)}
\nonumber\\
&&\nonumber\\
& = & \lim_{\eta\to 0+}\frac{1}{\sqrt{\pi t}}\,
\int_{-\infty}^\infty ds\ e^{-s^2/t}
\,\frac{1}{2}\partial_s\left(\frac{1}{\sin(2s+i\eta)}\right)
\nonumber\\
&&\nonumber\\
& = & \lim_{\eta\to 0+}\frac{1}{2t\sqrt{\pi t}}\,
\int_{-\infty}^\infty ds\ \frac{2s}{\sin(2s+i\eta)}\,e^{-s^2/t}\quad.
\label{eq:d4}
\end{eqnarray}
In the limit $\eta\to 0+$, the integrand has $\delta$--function 
contributions at $s=\frac{\pi}{2}\,n$, where $n\neq 0$ is an integer. 
These lead to the non--analytic corrections $\sim e^{-\pi^2 n^2/(4t)}$, 
which we will neglect in the following. The analytic corrections arise 
from small $s$--contributions in the integral. Expanding $(2s)/\sin(2s)$ 
in a power series (formula 4.3.68 of Ref. \cite{abr}) and evaluating the 
Gaussian integrals we arrive at:
\begin{equation}
\label{eq:d5}
[1]_m^{(t)}=\frac{1}{2t}\sum_{m=0}^\infty \left[
\frac{(-1)^{m-1} 2(2^{2m-1}-1)}{m!}B_{2m}\right] t^m
+{\cal O}(e^{-\pi^2/(4t)})\quad.
\end{equation}
Remarkably, the coefficients differ only by the sign $(-1)^m$ 
from those in (\ref{eq:d3}). Again, the first three terms yield 
(\ref{eq:57}).

%%%% page 23
\section{Calcuation of the coefficients (\ref{eq:67})}

In order to expand $S_1(k,\Delta t)$ we exploit the fact
that the $\lambda$--integration 
in (\ref{eq:49}) can be viewed as the formal application of the 
operator $\exp[-D(\mu)\Delta t]$ on $(1+\mu)^{-1}$ (with $D(\mu)$ as 
in (\ref{eq:7})):
\begin{eqnarray}
\int_0^\infty d\lambda\ g(\mu,\lambda;\Delta t)\frac{1}{1+\lambda} 
& = & e^{-D(\mu)\Delta t}\left(\frac{1}{1+\mu}\right) = 
\sum_{n=0}^\infty \frac{(\Delta t)^n}{n!}[-D(\mu)]^n
\left(\frac{1}{1+\mu}\right)
\nonumber\\
&&\nonumber\\
& = & \sum_{n=0}^\infty \frac{(\Delta t)^n}{n!}\ R_n(\mu) \quad.
\label{eq:e1}
\end{eqnarray}
Here, $R_n(\mu)=[-D(\mu)]^n (1+\mu)^{-1}$ is a polynomial of degree 
$n+1$ in $(1+\mu)^{-1}$, i.e. we may write
\begin{equation}
\label{eq:e2}
R_n(\mu)=\sum_{l=0}^n r_{n,l}\,(1+\mu)^{-(l+1)} \quad,
\end{equation}
where the coefficient $r_{n,l}$ obeys the recurrence relation
\begin{equation}
\label{eq:e3}
r_{n+1,l}=-4 l^2\,r_{n,l-1}+(2l+1)^2\,r_{n,l}
\end{equation}
with $r_{0,k}=\delta_{0,k}$. To calculate the coefficients $a_n$, $b_n$, 
$c_n$ in (\ref{eq:66}), we need the values of $r_{n,n}$, $r_{n,n-1}$ and 
$r_{n,n-2}$. From (\ref{eq:e3}) we find
\begin{eqnarray}
r_{n,n} & = & (-4)^n\,n!^2\quad,
\nonumber\\
&&\nonumber\\
r_{n,n-1} & = & (-4)^{n-1}\,(n-1)!^2\sum_{\nu=0}^{n-1}(1+2\nu)^2\quad,
\label{eq:e4}\\
&&\nonumber\\
r_{n,n-2} & = & (-4)^{n-2}\,(n-2)!^2\sum_{\nu=0}^{n-2}(1+2\nu)^2
\sum_{\tilde \nu=0}^\nu (1+2\tilde \nu)^2\quad.
\nonumber\\
\end{eqnarray}
On the other hand, we may insert (\ref{eq:e1}) and (\ref{eq:e2}) in 
(\ref{eq:49}) and perform the $\mu$--integration using (\ref{eq:39}), giving
\begin{equation}
\label{eq:e5}
S_1(k,\Delta t)=\sum_{n=0}^\infty \frac{(\Delta t)^n}{n!}
\sum_{l=0}^n r_{n,l}\ \frac{1}{(l+1)!^2}
\prod_{\nu=0}^l \left({\textstyle (\frac{1}{2}+\nu-i\frac{k}{2})\,(\frac{1}{2}+
\nu+i\frac{k}{2})}\right)\quad.
\end{equation}
%%%% page 24
The product is a polynomial of degree $(l+1)$ in $k^2$, of which we 
need the first three terms:
\begin{eqnarray}
\prod_{\nu=0}^l \left({\textstyle (\frac{1}{2}+\nu-i\frac{k}{2})\,(\frac{1}{2}+
\nu+i\frac{k}{2})}\right) & = & \frac{1}{4^{l+1}}
\prod_{\nu=0}^l \left(k^2+(1+2\nu)^2\right)
\nonumber\\
&&\nonumber\\
& = &\frac{1}{4^{l+1}}\left((k^2)^{l+1}+A_l\,(k^2)^l+ B_l\,(k^2)^{l-1}
+\cdots\right)
\label{eq:e6}
\end{eqnarray}
with
\begin{equation}
\label{eq:e7}
A_l=\sum_{\nu=0}^l (1+2\nu)^2\quad,\quad B_l=
\sum_{0\le \nu < \tilde \nu \le l} (1+2\nu)^2\,1+2\tilde\nu)^2\quad.
\end{equation}
Together, (\ref{eq:e5}) and (\ref{eq:e6}) imply
\begin{eqnarray}
a_n & = & \frac{(-1)^{n+1}}{4^{n+1}(n+1)!^2}\,r_{n,n}\quad,
\nonumber\\
&&\nonumber\\
b_n & = & \frac{(-1)^{n+1}}{4^{n+1}(n+1)!^2}\,r_{n,n}\, A_n+
\frac{(-1)^{n}}{4^n n!^2}\,r_{n,n-1}\quad,
\label{eq:e9}\\
&&\nonumber\\
c_n & = & \frac{(-1)^{n+1}}{4^{n+1}(n+1)!^2}\,r_{n,n}\, B_n+
\frac{(-1)^{n}}{4^n n!^2}\,r_{n,n-1}\,A_n
 +\frac{(-1)^{n-1}}{4^{n-1} (n-1)!^2}\,r_{n,n-2}\quad. 
\nonumber
\end{eqnarray}
for the desired coefficients.
Combining (\ref{eq:e4}), (\ref{eq:e7}) and (\ref{eq:e9}), one indeed 
obtains after some (computer--)algebra the result (\ref{eq:67}). It 
remains to show that $S_2(m,\Delta t)$ is given by (\ref{eq:65}) with 
the {\it same} coefficients (\ref{eq:67}). Expanding 
(\ref{eq:49a}) we find:
\begin{equation}
\label{eq:e10}
S_2(m,\Delta t)=\sum_{n=0}^\infty \frac{(-\Delta t)^n}{n!}\,(-2)
\,C_n(m)
\end{equation}
with
\begin{equation}
\label{eq:11}
C_n(m)=\sum_{\nu=1}^m \frac{1}{\nu}\sum_{l=0}^{\nu-1} (1+2l)^{1+2n}\quad.
\end{equation}
For small $n$ the $C_n(m)$ can be calculated with the help of
standard mathematical formulas. For our purposes, however, we need
the large--$m$ behavior for all $n$. Therefore we consider the
generating function
\begin{equation}
\label{eq:e12}
g(m,x)=\sum_{\nu=1}^m \frac{1}{\nu}\sum_{l=0}^{\nu-1}
\sinh[(1+2l)x]=\sum_{n=0}^\infty \frac{x^{1+2n}}{(1+2n)!}\,C_n(m)\quad.
\end{equation}
Using (twice) the finite geometric series we can derive the 
inhomogeneous differential equation
\begin{equation}
\label{eq:e13}
\sinh(2x)\,g(m,x)+(\cosh(2x)-1)\,\partial_x g(m,x)=
\cosh[(1+2m)x]-\cosh x,
\end{equation}
from which we get the recurrence relation
\begin{equation}
\label{eq:e14}
C_n(m)=\frac{1}{b_{nn}}\left(\eps_m^{n+1}-1-\sum_{k=0}^{n-1}
b_{nk}\,C_k(m)\right)
\end{equation}
with
\begin{equation}
\label{eq:e15}
b_{nk}=\frac{n+k+2}{2n+3}{2n+3 \choose 2k+1}\,2^{2(n-k+1)}\quad.
\end{equation}
Eq. (\ref{eq:e14}) indeed gives after some algebra
\begin{equation}
\label{eq:e16}
(-2)C_n(m)=a_n\eps_m^{n+1}+b_n\eps_m^{n}+c_n\eps_m^{n-1}+\cdots
\end{equation}
with $a_n$, $b_n$, $c_n$ as in (\ref{eq:67}). 

We conclude this appendix with a short remark: Equations 
(\ref{eq:64}) and (\ref{eq:65}) suggest that the functions
$S_1(k,\Delta t)$ and $S_2(m,\Delta t)$ are connected
by the analytic replacement $-k^2\to \eps_m=(1+2m)^2$ and 
$\Delta t\to -\Delta t$. We have verified this for the lowest
($n\le 2$) polynomials $r_{n+1}(z)$ and --- as far as the
highest coefficients $a_n$, $b_n$, $c_n$ appearing in (\ref{eq:66})
are concerned --- for all $n$. This is sufficient for the purpose
of the present paper. A general proof, however, has yet to be given.

%%%%%%%%%%%%%%%% References %%%%%%%%%%%%%%

%\end{multicols}

\end{document}